\def\gtsim{\raise 2pt \hbox {$>$} \kern-1.1em \lower 4pt \hbox {$\sim$}}
\def\ltsim{\raise 2pt \hbox {$<$} \kern-1.1em \lower 4pt \hbox {$\sim$}}

\def\farcs{\hbox{$.\!\!^{\prime\prime}$}}
\def\fmag{\hbox{$.\!\!^{\rm m}$}}

\documentclass[10pt, twocolumn]{article}
\usepackage{graphics}
\usepackage{rotate}
\skip\footins 10pt plus 2pt minus 4pt

\begin{document}
\title{Particle Accelerators in the Hot
Spots of the Radio Galaxy 3C\,445, Imaged With the VLT}

\author{M. Almudena Prieto$^{\ast}$, Gianfranco Brunetti${\dag}$,
Karl-Heinz Mack${\ddag}$\\
$^{\ast}$European Southern Observatory, Karl-Schwarzschild-Str. 2,
D-85748
Garching, Germany\\
 email: aprieto@eso.org\\
 ${\dag}$Istituto di Radioastronomia del CNR, Via P. Gobetti 101,
 I-40129 Bologna, Italy\\
 email: brunetti@ira.cnr.it\\
 ${\ddag}$ASTRON/NFRA, Postbus 2, NL-7990 AA Dwingeloo, The
Netherlands\\
Istituto di Radioastronomia del CNR, Via P. Gobetti 101,
I-40129 Bologna, Italy\\
Radioastronomisches Institut der Universit\"at Bonn, Auf dem H\"ugel 71,
D-53121 Bonn, Germany\\
email: mack@ira.cnr.it
}

\maketitle

{\bf
Hot spots (HSs) are regions of enhanced radio emission  produced
by supersonic jets at the tip of the radio lobes of  powerful radio
sources. Obtained with the Very Large Telescope (VLT), images
of the HSs in the radio galaxy 3C\,445 show bright knots
embedded in diffuse optical emission  distributed along
the post shock region created by the impact of the jet into the
intergalactic medium.

The observations reported here confirm that relativistic electrons are
accelerated by Fermi-I acceleration processes in HSs.
Furthermore, both the diffuse emission tracing the rims of the front shock
and the multiple knots demonstrate the presence of additional
continuous re-acceleration processes of electrons (Fermi-II).}

HSs are common features in powerful extragalactic lobe-dominated radio
sources (Fanaroff-Riley class II). Their radio spectra are
interpreted in terms of synchrotron radiation. Detection of optical
radiation from HSs
is difficult due to the natural decay of the synchrotron
spectrum toward higher frequencies. Yet, the optical detection of HSs
has
a number of fundamental implications in our understanding of the
physics of particle acceleration in these regions.
Relativistic electrons rapidly lose their energy-emitting synchrotron
radiation and the radiative lifetime of the electrons
responsible for the emission in the optical band is
$\sim 300$ times shorter  -- between a few hundred to a few thousand
years, depending on the magnetic field strength -- than that of the
less energetic electrons emitting in radio wavelengths.
As a consequence,
optical detection of HSs implies either an extremely
efficient transport of the ultra-relativistic
particles from the core to the HS ({\it 1}),
or the in-situ production  of such
energetic particles with Lorentz factors $ \gamma >10^5$ to $10^6$
, where the Lorentz factor $\gamma$ is defined as
$\gamma = (1-(\frac{v}{c})^2)^{-\frac{1}{2}}$ ({\it 2}).
Although the search for optical emission from radio HSs has a
long history ({\it 3 - 5}), few HSs have been detected
as synchrotron emitters at optical frequencies ({\it 6}).
Indeed, at high frequencies most well studied HSs show an abrupt
steepening of
the spectrum below the straight power law
extrapolated from the low frequencies, which is interpreted as a
result of a high energy cut-off in the spectrum of the synchrotron
emitting electrons.
Thus far, there are only two known HSs in which the radio
power law spectrum can be extrapolated up to the optical or ultraviolet
regime (3C\,303, {\it 7}) or even to the X-ray regime (3C\,390.3, {\it 8}),
indicating
the continuation of the synchrotron emission to  higher energies.
In the case of optical detection, the short radiative lifetime of
the emitting electrons as well as their diffusion length scale
of much less than 1 kpc make high-resolution optical imaging (so far
obtained only with Hubble Space Telescope) {\it 9} an unequaled tool with 
which to sample those regions where high energy electrons are injected
and/or re-accelerated.

Here, we report on the imaging of the
local accelerator regions in the southern HS of the radio galaxy 3C\,445.
3C\,445 is a classical double-lobe radio source at redshift z=0.0562.
At the tips of each lobe are the southern HS, placed at $\sim 450$ kpc
(${\rm H}_{\rm 0} = 50~ {\rm km}~ {\rm s}^{-1} {\rm Mpc}^{-1}$,
${\rm q}_{\rm 0} = 0.5$), and the northern HS at a
projected distance of $\sim 460$ kpc from the core ({\it 10}).
Both HSs were observed with the Very Large Telescope (VLT)
of the European Southern Observatory and the Infrared Spectrometer and Array
Camera, at the bands K$_{\rm s}$ (2.2 $\mu$m), H (1.7 $\mu$m) and J$_{\rm s} $
(1.2 $\mu$m), and the Focal Reducer/low dispersion Spectrograph,
at the I-band (0.9 $\mu$m).
Each final image is the result of stacking between
10 (I-band) and 30 to 40 (J$_{\rm s}$-, H- and
K$_{\rm s}$-band) slightly shifted frames.
Sky background removal was done by a median filter of several adjacent frames.
The integration time ranged from 35 to 60 min depending on the band. The
final image quality has an equivalent full width at half maximum of 0\farcs6
in K$_{\rm s}$, 0\farcs5 in H,
0\farcs45 in J$_{\rm s}$, and 0\farcs55 in I.
The counterpart of the radio emission from each HS was detected in all
bands. The emission is extended in both cases and in all filters,
but only the emission in the southern HS  could be resolved in several
compact regions.  Thus, we focus our analysis and discussion on
the southern HS.

The emission from  the southern HS of 3C\,445 is resolved in at
least three compact regions (Fig.~1). These regions are further
enclosed by diffuse emission detected at about the 5$\sigma$ level in all
bands; the diffuse emission is more evident in the lower-frequency images.
The overall optical emission has an arc-like shape and size as in the radio
image, presenting the typical  structure of a post-shock region usually
seen at radio waves ({\it 10}).
We checked whether the electrons responsible for the
synchrotron emission could have been transported by the jet
from the core of the galaxy to the HS.
The maximum distance electrons can travel away from the galaxy nucleus
depends on both their synchrotron and inverse Compton (IC) radiative
lifetimes ($\propto 1/\gamma$) as well as on the velocity of the jet.
The Lorentz factor of the electrons emitting synchrotron
radiation in the I-band is $\gamma \sim 3\times10^6 B^{-1/2}$,
where $B$ is the magnetic field strength in the HS region in nT.
A first estimate of $B$ can be derived assuming the usual equipartition
conditions ({\it 11}): From the observed synchrotron luminosity (Fig.~2)
and the radio HS size (4\farcs8$\times2''$), $B \simeq 2.8$ nT
and the corresponding Lorentz factor is thus $\gamma~ \sim 2\times 10^6$.
The maximum distance (projected on the plane of the sky) the optical electrons
in 3C\,445 South could have reached away from the nucleus of the
galaxy is less than 100 kpc, independent of the jet velocity and assuming
that the electrons suffer only IC losses due to the
interaction with the microwave photons (i.e. the intrinsic magnetic field is
neglected, {\it 12})
Yet, HS 3C\,445 South is 450 kpc from the core and, thus, the
optical electrons witnessed must be produced in place.

An effective mechanism to accelerate relativistic electrons in the HSs of
radio galaxies to the required energies,
$\gamma \sim 10^5$ to $10^6$, to emit optical light, is strong shocks.
Strong shocks are expected to be produced by the impact of the jet into the 
intergalactic medium ({\it 6,13}).
 From an observational point of view, the bow-shock shape seen in the
 images of the southern HS in 3C\,445, at the tip of the radio lobe supports
the shock-wave scenario.
The integrated magnitudes in the K$_{\rm s}$- ($18\fmag5\pm0\fmag09$),
H-($19\fmag5\pm0\fmag2$),
J$_{\rm s}$- ($20\fmag3\pm0\fmag1$)
and I-bands ($21\fmag3\pm0\fmag1$) of
the HS, together with the integrated emission at 8.4 GHz
($90 {\rm mJy} \pm 9$ mJy), were fit to theoretical synchrotron
spectra produced by a population of accelerated electrons
[after ({\it 14})].
Assuming a power-law electron distribution with
typical injection spectral indices $\delta = 2.0$ to 2.2,
the models can successfully reproduce the data with a best fit cut-off
frequency ($\nu_c$, the synchrotron frequency emitted
by the electrons accelerated at the maximum energy),
$\nu_c \sim 1 - 2 \times 10^{15}$ Hz (Fig.~2). For those parameters,
the break of the synchrotron spectrum ($\nu_b$, the synchrotron frequency
emitted by electrons of Lorentz factor $\gamma_b$ beyond that radiative
losses cause a steepening of the electron spectrum)
occurs at $\nu_c/\nu_b \simeq 300\pm 100$ for $\delta =2.2$, and at
$\nu_c/\nu_b \simeq 1000 \pm 300$ for $\delta =2.0 $ (errors at 90\%
confidence level).
Because there are no radio observations with sufficiently high
resolution, attempts to determine individual synchrotron spectra
for the brightest compact regions  in the HS are currently inconclusive;
however  $\nu_{\rm c} > 3 \times 10^{14}$ Hz is obtained for the three main
knots in the HS.

Numerical simulations of HSs ({\it 15 - 18}) show that a high
pressure cap is formed at the tip of the jet, from where the pressured plasma
escapes backward into a larger, diffuse region, or cocoon, that forms
around the jet.
The VLT observations show  a `primary'  knot,
identified as the tip of the jet characterized by its enhanced
brightness; two `secondary' fainter knots, on each side
of the primary; and diffuse  emission with a bow-shock shape
that we associate with the emission from the plasma which escapes from the
shock region into the cocoon.
However, the detection of emission up to 3 to 5 kpc away from the primary knot,
is surprising because, assuming an equipartition magnetic field
in the HSn the optical emitting electrons have a radiative lifetime
of $\sim 2\times 10^4$ yr
and thus a typical travel length $\ll 1$ kpc.
Consequently, unless an improbably low intrinsic magnetic field
$B < 0.5\;$nT
is assumed,  optical and infrared emission on scales larger than 1 kpc
provides direct evidence for additional in-situ re-acceleration in the
HS region with an acceleration time scale of the order of $10^4$ yr.
Similar evidence has recently been observed in extragalactic optical jets
(e.g., 3C\,273, {\it 19}).

In addition, the appearance of multiple knots may
indicate the presence of incipient Rayleigh-Taylor instabilities that produce
variations in the plasma density and path length,
leading to multiple emitting components such 
as those expected by simulated synchrotron images of HSs ({\it 18,20}).
The development of Rayleigh-Taylor instabilities in the cocoon
may power magnetohydrodynamic (MHD) turbulence that can account for the
required additional re-acceleration of the relativistic electrons
downstream.
Assuming Alfv\'{e}nic turbulence in the cocoon, the MHD Fermi-II-like
acceleration time {({\it 21}) and
ref. therein] is :

\begin{equation}
\tau^{II}_{\rm acc} ({\rm sec}) \sim {{ l
\, c }\over{ v_{\rm A}^2 }}
\left( {{\delta B}\over{B}} \right)^{-2}
\end{equation}

where $l$ is the typical distance between
the peaks of turbulence responsible for Fermi-II acceleration,
$v_{\rm A}$ is the Alfv\'{e}n velocity, and c is the speed of light.
In following with Eq.~1, the  acceleration efficiency $(1/\tau)$,
is proportional to the energy density in turbulence waves with scales $< l$,
 defined as $(\delta B)^2/8\pi$.
The scale $l$ should be between the minimum scale
($\sim 2\pi v_{\rm i} / \Omega_{\rm i}$, $v_{\rm i}$ and
$\Omega_{\rm i}$ being the sound velocity and the cyclotron
frequency of the ions, respectively) and the maximum scale of the turbulence
in the plasma.  Two regimes of plasma temperatures are considered.
For $T < 10^7$ K, the largest possible scale of the undamped Alfv\'{e}nic
turbulence is $l_{\rm max}= 2 \pi /k_{\rm min}$,
where $k_{\rm min} \sim (\epsilon/ \nu^3_{\rm D})^{1/4}$ m$^{-1}$
[({\it 22}) and references therein].
The dissipation coefficient $\nu_{\rm D}$ relevant for the damping of the
turbulence waves
is $\nu_{\rm D} \sim 3.8 \times 10^{10} T^{5/2}/n_{\rm p}$, whereas
$\epsilon$ (J kg$^{-1}$ s$^{-1}$) is the energy source of the
fluctuations that balances the dissipation.
For proton thermal densities $n_{\rm p}\sim 10^{2}$ m$^{-3}$ assuming
$\epsilon \sim 0.1$ J kg$^{-1}$ s$^{-1}$
typical of HSs ({\it 22}), the maximum scale of the turbulence is found
$l_{\rm max} \sim 20(T/5\times 10^5)^{15/8}$ pc. Substituting this
value in Eq.~1, with $\delta B/B \sim 1$ and
$v_{\rm A} \sim 5 \times 10^6$ m s$^{-1}$,
the Fermi-II-like acceleration time is
\begin{equation}
\tau_{\rm acc}^{II} \sim 2\times 10^{5} (l/l_{\rm max}) (T/5 \times 10^5)^{15/8}\;\; {\rm yr}.
\end{equation}
As a consequence, waves with
turbulence scales of $\sim 0.1 \times l_{\rm max}$ (which account for
the bulk of the energy) can easily power Fermi-II processes.
For $T > 10^7$ K, which is typical of
the intracluster medium, damping processes are less
efficient and the largest turbulence scale is limited
by the injection scale: here, of the order
of the transverse dimension of the downstream region,
i.e., $l_{\rm max} \sim 1$ kpc.
Assuming $\delta {B_k}^2 \propto k^{-s}$
($\delta {B_k}^2/8\pi$ is the differential energy density in waves
between k and k+dk and $s=5/3$ in the Kolmogorov case,
$s=3/2$ in the  Kraichnan case)
and $\delta B^2 = \int dk \delta {B_k}^2 \sim B^2$,
we find that in the Kraichnan case,
$\tau_{\rm acc}^{II} \sim 3 - 6 \times 10^4$ yr for
$l \sim 0.1 - 0.2$ pc. Waves with that turbulence scale contain
enough energy density to accelerate the optical emitting electrons.
However, in the Kolmogorov case MHD Fermi-II-like
acceleration is less efficient and very strong turbulence,
$\delta B /B \sim 2 - 5$, is required.

To test whether additional Fermi-II-like
acceleration in the downstream region can account for
the observed properties of the observed HS,
 a calculation in three phases was  considered.
This calculation attempts to  mimic the behavior of the plasma in the HSs as
seen in MHD simulations.
First, the relativistic electrons diffuse, expanding from the high pressure
region (central primary knot), being also subject to synchrotron and IC losses.
Second, an approximate pressure balance is reached and electrons suffer only
synchrotron and IC losses. Finally, the plasma is recompressed in the regions
of the two secondary knots, possibly due to the development of non-linear
plasma instabilities.
During the three phases,  relatively efficient ($\tau_{\rm acc}^{II}
\sim 2 - 5 \times 10^4$ yr) second-order Fermi re-acceleration
mechanisms are assumed.
We calculated the evolution of the electron energy considering
a magnetic field frozen into the plasma (homogeneous) and
obtained the synchrotron cut-off frequency, $\nu_{\rm c}$, as
a function of the distance from the primary knot for different model
parameters (Fig.~3).
Assuming a canonical transport velocity from the primary knot
in the range 0.03 to 0.1\,c, we found that
the combined effect of adiabatic expansion (during the first phase)
and second-order Fermi
processes can maintain $\nu_{\rm c}$ high enough to have optical and infrared
emission from
secondary knots located at $\sim$ 3 kpc distance from the primary
knot and still have detectable residual extended optical and infrared 
emission in between the knots.
More specifically, we find that
this scenario works only if we assume relatively strong expansion
factors during the first phase ($f > 2$, $f$ is defined as the
ratio between the size of the plasma region after and before the
compression or expansion), to alleviate the strong synchrotron
losses in the region of the primary knot, and if we assume 
a relatively low initial magnetic field strength $B \leq 15\;$nT (at
least for a  conservative $f > 0.25$ during the third phase).
We also found that such a limit
yields a magnetic field strength averaged over the total
HS region $B \ltsim 4 nT$, which is consistent with the
field strength estimated from equipartition.
The above limit on the magnetic field strength and the derived
$\nu_{\rm c}$  from the synchrotron spectrum allow constraining  the
efficiency of the acceleration in the primary knot, where the
relativistic electrons are accelerated directly by the
shock (Fermi-I).
The acceleration efficiency here is higher:
$\tau_{\rm acc}^{I} \sim 10^3 \left( B /15
\right)^{-3/2}$year, as expected.\\\\

{\flushleft \Large \bf References and Notes}
\begin{itemize}
\item[1.] W. Kundt,  Gopal-Krishna, Nature, {\bf 288}, 149, 1980
\item[2.] K. Meisenheimer {\it et al.}, A\&A {\bf 219}, 63, 1989
\item[3.] W.C. Saslaw, J.A. Tyson, P. Crane, ApJ {\bf 222}, 435, 1978
\item[4.] S.M. Simkin, ApJL {\bf 222}, 55, 1978
\item[5.] P. Crane, J.A. Tyson, W.C. Saslaw, ApJ {\bf 265}, 681, 1983
\item[6.] K. Meisenheimer, M.G. Yates, H.-J. R\"oser, A\&A {\bf 325},
57, 1997
\item[7.] W.C. Keel, ApJ {\bf 329}, 532, 1988
\item[8.] M.A. Prieto, MNRAS {\bf 284}, 627, 1997
\item[9.] R.C. Thomson, P. Crane, C. Mackay, ApJL {\bf 446}, 93, 1995
\item[10.] J.P. Leahy {\it et al.}, MNRAS {\bf 291}, 20,
1997
\item[11.] A.G. Pacholczyk, Freeman \& Co., San Francisco, 1970
\item[12.] Gopal-Krishna, P. Subramanian, P.J. Wiita, P.A. Becker,
A\&A {\bf 377}, 827, 2001
\item[13.] G. Brunetti {\it et al.}, ApJL {\bf 561}, 157, 2001
\item[14.] G. Brunetti, M. Bondi, A. Comastri, G. Setti,
A\&A {\bf 381}, 795, 2002
\item[15.] M.L. Norman, K.-H.A. Winkler, L. Smarr, M.D. Smith,
A\&A {\bf 113}, 285, 1982
\item[16.] S. Massaglia, G. Bodo, A. Ferrari, A\&A
{\bf 307}, 997, 1996
\item[17.] J.M. Marti, E. M\"uller, J.A. Font, J.M. Ibanez, A. Marquina,
ApJ {\bf 479}, 151, 1997
\item[18.] M.-A. Aloy, J.-L. Gomez, J.-M. Ibanez, J.-M. Marti, E.
M\"uller,
ApJL {\bf 528}, 85, 2000
\item[19.] S. Jester, H.-J.
R\"oser, K. Meisenheimer, R. Perley, R. Conway, A\&A {\bf 373}, 447,
2001
\item[20.] M.D. Smith, M.L. Norman, K.-H.A. Winkler, L. Smarr, MNRAS
{\bf 214}, 67, 1985
\item[21.] M. Gitti, G. Brunetti, G. Setti, A\&A {\bf 386}, 456, 2002
\item[22.] C. Lacombe, A\&A {\bf 54}, 1, 1977
\item[23.] The positional coincidence
between optical, infrared  and radio data has been very
carefully checked using the galaxy core position (for the radio maps)
and
the position of nearby stars taken from the USNO-A2.0 catalogue (for the
optical maps, accuracy: 0\farcs25).
\item[24.] KHM was supported by a Marie-Curie-Fellowship
of the European Commission.  GB aknowledges partial
finantial support from the Italian MIUR (Ministero dell'Istruzione
dell'Universit\`a e della Ricerca) under grant Cofin01-02-8773.
We would like to thank the staff at the VLT Paranal Observatory for
their professional service.
We are grateful to M. Bondi, I.M. Gioia, M.J.M. March\~a, K. Meisenheimer,
M.-A. P\'erez-Torres, G. Setti and S.R. Spangler for their valuable comments
on the manuscript.

\end{itemize}

\newpage
\begin{figure*}[t]
\resizebox{170mm}{!}{\rotatebox{0}{\includegraphics*{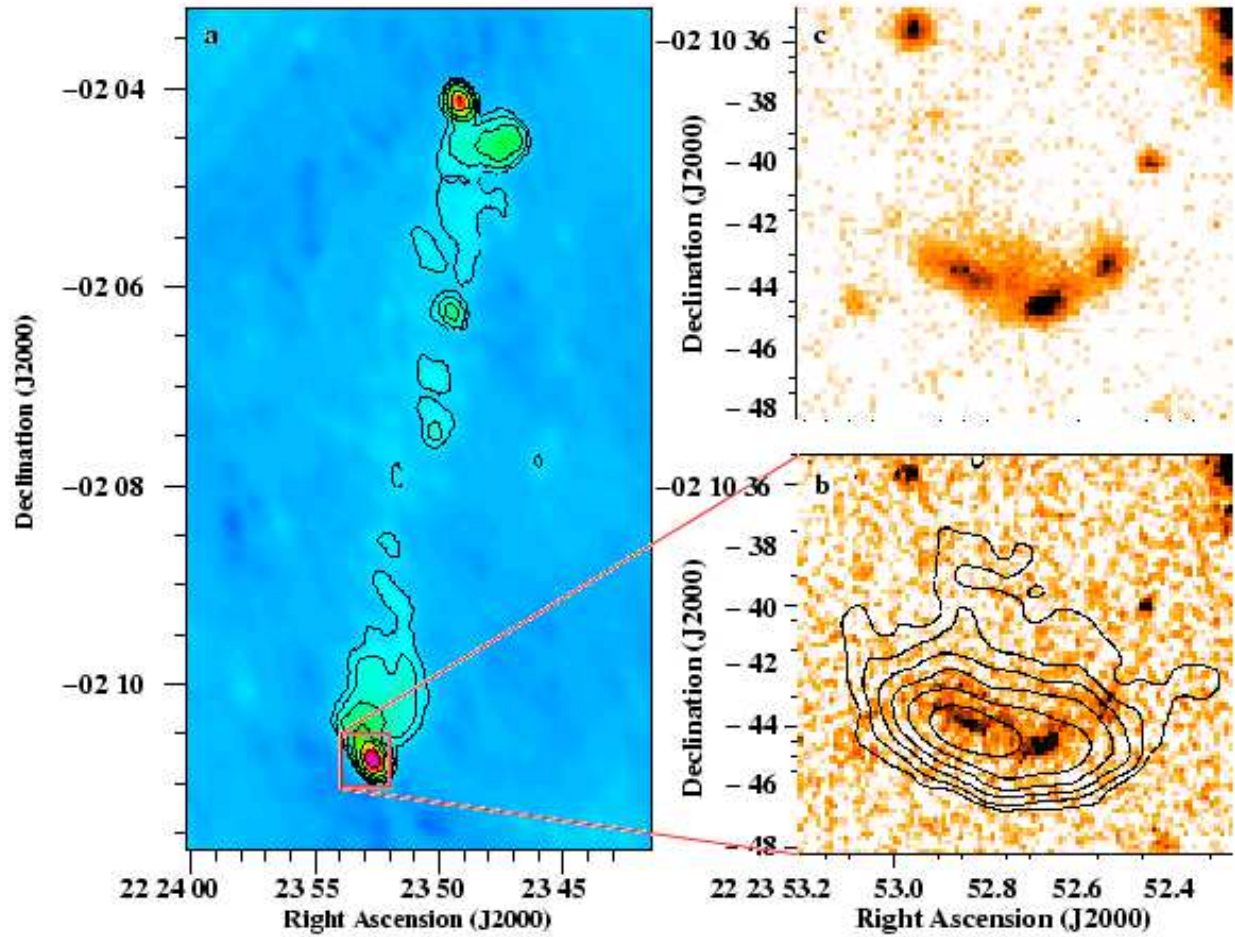}}}

\caption
{
{\bf (A)}:
Radio contours and colors from a snapshot observation
of 3C\,445 obtained with the Very Large Array (VLA) in D-configuration
at 8.4 GHz; contour levels are 2, 4, 8, 16, 32, 64 mJy/beam. Color wedge
scales from 0.5 mJy/beam (blue, background noise level) to 100 mJy/beam (red,
peak flux density).
{\bf (B)}:
Enlarged image of the southern HS region observed at
the same frequency with the VLA B-array; the half-power beam
width is 1\farcs1.
The radio contours show the southernmost part of the HS and are
superimposed
on the VLT J$_{\rm s}$-image ({\it 23}); the radio maps were
reduced from data requested from the National Radio Astronomy Observatory
VLA archive;
contour levels are 0.25, 0.5, 1, 2, 4, 8 mJy/beam; color wedge scales linearly
from $\sim 23.8$ (darkest emission), 25.4 (diffuse emission) to 25.5
(background) mag/beam.
{\bf (C)}:
The VLT I-band image of the southern HS region; color wedge ranges
from $\sim  24.5$ (darkest), 26.2 (diffuse) to 27 (background)
mag/beam.
}
\end{figure*}

\newpage
\begin{figure*}
\resizebox{15cm}{!}{\includegraphics*{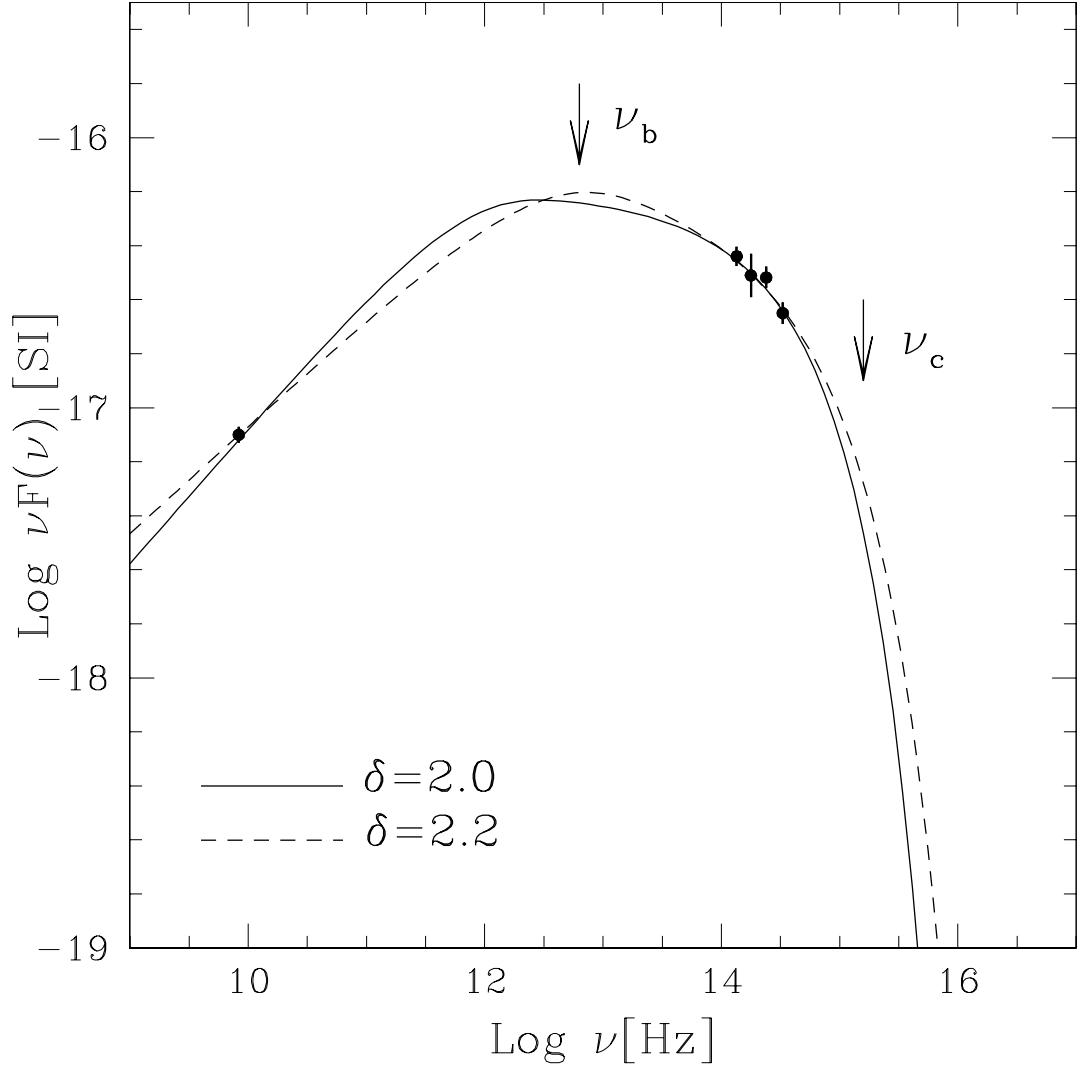}}
\caption{
Best synchrotron model fits to HS 3C\,445 South.
The solid circles denote the energies as obtained from the radio data
(8.4 GHz) and the infrared and optical measurements (K, H, J, and I band).
The models correspond to an injection spectral index of the
electron distribution $\delta =  2.0$ or $2.2$.
The best-fit cut-off and break frequencies are
$\nu_{\rm c} = 1.0 \times 10^{15}$ Hz and
$\nu_{\rm b} = 1.0 \times 10^{12}$ Hz ($\delta =  2.0$)
or
$\nu_{\rm c} = 1.8 \times 10^{15}$ Hz and
$\nu_{\rm b} = 5.9 \times 10^{12}$ Hz ($\delta =  2.2$), respectively.
The positions of the best-fit cut-off and break frequencies
in the case of $\delta=2.2$ are indicated by the arrows.
}
\end{figure*}

\newpage
\begin{figure*}
\resizebox{21cm}{!}{\includegraphics*{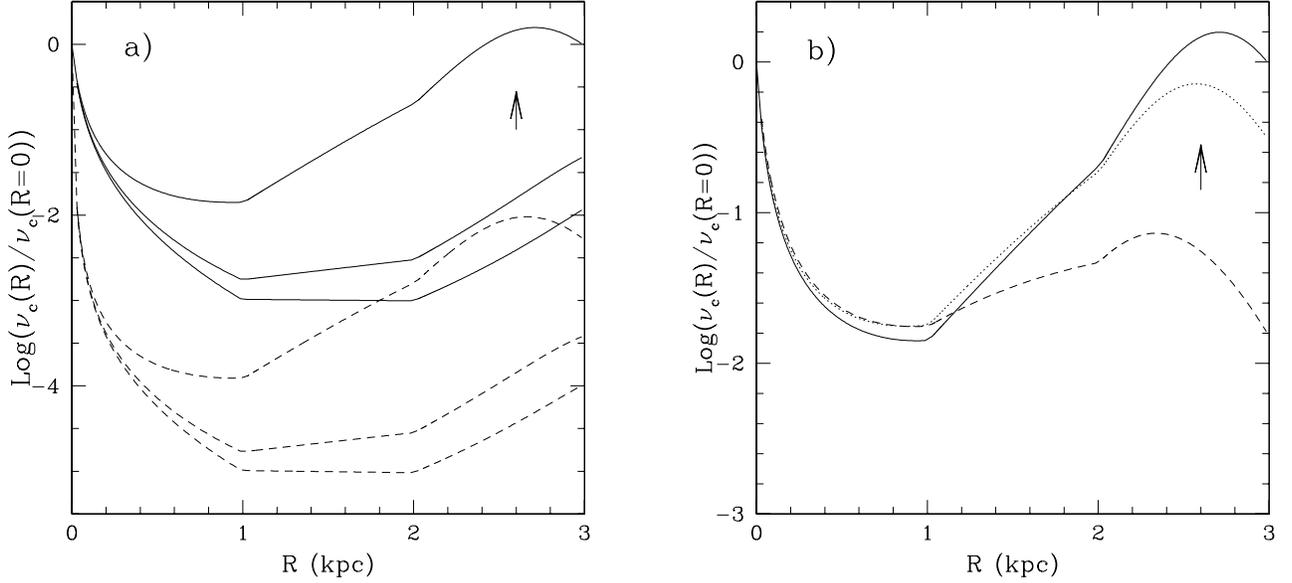}}
\caption{
The calculated cut-off frequency, $\nu_{\rm c}$, of the synchrotron
spectrum (relative to the value at the ``primary'' knot)
is plotted  as a function of the distance R from the primary knot
(at R=0); the farthest ``secondary'' knot is at R=3 kpc.
The arrows give the limits of $\nu_{\rm c}$ for the stronger secondary knot.
The calculations do not take into account the finite resolution obtained in
the VLT images (Fig.~1, spatial resolution $\sim 0.7$ kpc) and  the
projection effects.
The lengths and durations in the three-phase calculation are assumed to be
similar and adjusted to $l = 1$ kpc to match the observed distribution of the
knots.
{\bf (A)} model calculations for
an expansion factor $f=3$ during the first phase (0 to 1 kpc) and
a compression factor $f=0.5$ during the third phase (2 to 3 kpc),
assuming as initial magnetic field strengths
$B=8$ nT (solid lines) and $B=30$ nT (dashed lines).
 From the bottom to the top of the diagram, for each value of $B$,
the Fermi-II acceleration time is $\tau^{II}_{\rm acc}$=
$10^8$, $10^5$ and $2\times 10^4$ year, respectively.
{\bf (B)} model calculations for an initial magnetic field strength
$B = 8$ nT, for a Fermi-II acceleration time $\tau^{II}_{\rm
acc} = 2 \times 10^4$ yr and for a compression factor $f=0.5$ during the third
phase (2 to 3 kpc). Different lines correspond to different expansion factors
during the first phase (0 to 1 kpc): $f=3$ (solid line), $f=2.5$
(dotted line) and $f=1.5$ (dashed line).
}
\end{figure*}

\end{document}